# Analysis of Performance of Drivers and Usage of Overtime Hours: A Case Study of a Higher Educational Institution

**by KC Sanjeevani Perera**

## 1. Introduction

The University of Moratuwa has a vehicle fleet of 37 vehicles: 8 cars, 8 buses, 12 vans, 3 cabs 1 jeep, 1 truck, 2 hand tractors, and 2 three-wheelers. The University has an approved cadre of 38 driver posts and 27 permanent drivers are employed at the time of writing the paper.

The University drivers can be mainly categorized into two groups depending on the vehicle they are assigned to. The two categories are designated drivers of University Officers or drivers of assigned vehicles and drivers in the University vehicle pool. The scheme of recruitment of drivers specifies four grades of drivers, namely grade II, grade I supra grade, and special grade. The grade depends basically on the number of years they have served.

For all the categories of drivers, performance evaluation is done annually by the Officer in charge of the General Administration Division. The performance evaluation criteria officially used for the drivers is the same that is used to evaluate the staff doing desk jobs and cannot be considered as relevant to the job performed by the drivers.

The number of overtime hours allocated to a driver per month is 90 hours. However, all the drivers are allowed to work up to a maximum of 160 hrs of overtime per month as most of the vehicle reservations fall outside the normal working hours of a driver. In the cases of the designated drivers of the Vice-Chancellor and the Deputy Vice-Chancellor, a maximum of 180 hours of overtime per month is allowed as per the circulars.

It was observed that some drivers in the pool are working for a higher number of overtime hours per month than the other drivers which may be due to their high performance. This paper attempts to find out whether there is a correlation between the performance of the drivers and the number of overtime hours they have worked for. Also, to gain insight into the factors that affect the performance of drivers of the University and offer recommendations to improve the performance of low-performing drivers.

The next section of this paper comprises of review of literature on performance evaluations of employees, the qualities of drivers and evaluating the performance of drivers.

## 2. Literature Review

**Performance Evaluation (PE)**

Employee Performance Evaluation (EPE) is the Human Resource activity by means of which the organization determines the extent to which the employee is performing the job effectively. (Glueck, 1979 and Ivancevich, 1998). Performance appraisal was first used by the United States Army in World War One to assess the performance of the Officers. (Scott *et al.*, 1941)

Any employee performance evaluation system must have a formality and clear purposes for which it is used. Douglas McGregor[1] identifies that there are three basic purposes of conducting PE i.e. letting people know where they stand, identifying an individual's training and development needs and providing accurate performance data.

Opatha (2003) outlines the objectives of performance evaluations as threefold. They are as follows.

*Administrative Objectives* i.e. to grant salary increment, to manage promotions, transfers, rewards, to make decisions on probation period and for disciplinary management.

*Developmental Objectives* i.e. to develop the employee, to identify training needs, to improve productivity, to restructure the jobs and improve relationship between the superiors and the subordinates etc.

*Informative Purposes* i.e. to let the employee understand what is expected of him and also to help the employee to progress through the career.

James W. Smither (1998) describes performance evaluation as a process of identifying, observing, measuring and developing human productivity in organizations. He identifies eight characteristics of a successful performance evaluations system i.e. 1) make sure that the appraisal criteria are relevant to the job, 2) make sure appraisal criteria are clearly defined, 3) train raters on the appraisal process, 4) conduct appraisals frequently and allow enough time for raters to appraise ratees thoroughly, 5) make sure appraisals are appropriate for individual or team goals, 6) avoid overall appraisals, 7) use more than one rater if possible and 8) make raters accountable for their appraisals.

---

[1] From the book 'An uneasy look at Performance Appraisal'

Performance appraisal has three basic functions: (1) to provide adequate feedback to each person on his or her performance; (2) to serve as a basis for modifying or changing behavior toward more effective working habits; and (3) to provide data to managers with which they may judge future job assignments and compensation. The performance appraisal concept is central to effective management. Much hard and imaginative work has gone into developing and refining it. In fact, there is a great deal of evidence to indicate how useful and effective performance appraisal is. Yet present systems of performance appraisal do not serve any of these functions well with respect to the professional drivers.

In order to set performance criteria relevant to a job done by an employee, it is necessary to understand the qualities required to perform the job. In the case of drives, the qualities listed by *Traffic Counsel*, a US-based Traffic Law website are given below.

**Skilled -** One of the most important qualities of a good driver is that he or she should be skilled enough to handle any situation on the road. A good driver should have the skills to control the vehicle in every situation.

**Knowledge** - Skills are not the only quality that a good driver possesses. He or she should also have proper and complete knowledge about all the necessary rules of the road. A good driver also has knowledge about his or her vehicle.

**Self-Discipline** - It is very easy to advise other people to drive as per rules but being in self-discipline while driving is one of the best qualities of a good driver. Maybe crossing a red light makes you look cool for some time but also puts your life in danger while on road.

**Patience** -Whenever you are on the road, it is not a race who will get home or to work first, because everyone has a different destination. Hence, being patient is the best and necessary quality of a good driver.

**Alertness** - Not everyone in the road is not a good driver, there are many people on the road who are not so. Hence, to be a good driver, one must be absolutely alert all the time on the road for the safety of oneself and other people.

**Mechanical Skills** –a good driver should know the basic mechanical skills so that when the need arises, he or she can troubleshoot the basic issues of his or her vehicle. Any professional driver must have the understanding of modern vehicle dynamics.

**Responsibility** - One of the most important qualities of a good driver is that he or she is responsible. You will not always be traveling alone in the vehicle. Hence, it is very important to be responsible and take care of the people traveling with you.

**Care for vehicle** - Being a good driver also involves taking care of the vehicle one is driving. This means keeping a check on the timely servicing of the vehicle, ensuring reduction of fuel consumption and vehicle wear.

**Fitness** – Fitness is also one of the important qualities of a good driver. If a person is fit to drive, then only he or she is provided a license to drive a vehicle, no matter which vehicle it is.

**Customer service** - Aptitude for customer service is also an essential quality of professional drivers. Having necessary social etiquettes is a must for any professional driver. They must also have knowledge of local geography along with short cuts to save time. They must also be able to plan a route for comfort, efficiency and safety. They must be able to cope with a wide range of road surfaces and weather conditions. They must be aware about both personal and public safety measures to ensure the safety of the passengers.

**Problem-solving** - A good driver anticipates what other drivers are going to do and is aware of what is going on around them since drivers usually work independently, they are often responsible for resolving any problems that may occur on the job. If a road is closed or their vehicle has a flat tire, they must decide how to solve the issue quickly and logically.

**Concentration** - A driver needs to have a strong concentration when transporting passengers. You should be attentive and alert in case unexpected hazards are found on the road or to avoid any other cars from crashing into the vehicle. The eyes of the driver should be on the road at all times and ready for any obstacles. This can be done by keeping oneself away from any distractions while driving.

## 3. Study Design

This study followed the evaluative framework given below. The drivers were rated by the users on a five-point scale on the traits: skill, knowledge, patience, alertness, mechanical knowledge, care for the vehicle, customer service, responsibility, fitness, problem solving, and concentration. A random sample of 20 frequent users of university vehicles was picked for the study to evaluate the performance of drivers. The checklist used for the purpose of collecting data is given in the appendix.

Secondary data were collected from the overtime records of the drivers for the years 2017 and 2018. The overtime records of the years 2019, 2020 and 2021 were not used as the said three years do not reflect the actual usage of vehicle of the University due to restricted movements caused by Easter attacks in 2019 and COVID -19 pandemic in 2020 and 2021.

## Data Analysis and Discussion

The total number of drivers in the University is 27 and out of them, 8 serve as designated drivers of University Officers. The said 8 drivers were excluded from this study as the number of overtime hours worked by them depends mainly on the time taken to travel between the residence of the Officer and the University and the number of times the vehicle is used by the Officer to attend any official duties outside the University. Therefore, the actual number of drivers subjected to this study was 21.

Each driver in the pool was given a number to conceal the identity since it is not ethical to publish the number of overtime hours worked by each of them. The drivers who worked throughout the year were considered for this survey. Therefore, newcomers and those who got transferred to another University were not considered in the evaluation.

The type of vehicle and the condition of each vehicle assigned to each driver and the number of average over time hours worked by them per month are as follows.

| Driver | Type of Vehicle | Condition of Vehicle | Ave OT hrs per month 2018 | Ave OT hrs per month 2017 | Grand Average for two years |
|---|---|---|---|---|---|
| driver 1 | Van | Good | 117.83 | 134.23 | 126.03 |
| driver 2 | Minibus | Satisfactory (Non AC) | 86.75 | 77.77 | 82.26 |
| driver 3 | Bus | Good | 110.25 | 132.21 | 121.23 |
| driver 4 | Van | Satisfactory | 121.33 | 142.40 | 131.86 |
| driver 5 | Van | Very good | 132.50 | 149.69 | 141.09 |
| driver 6 | Van | Very good | 137.21 | 157.44 | 147.32 |
| driver 7 | Van | Very good | 124.71 | 129.05 | 126.88 |
| driver 8 | Bus | Good | 137.08 | 134.50 | 135.79 |
| driver 9 | Cab | Satisfactory | 102.98 | 94.77 | 98.875 |
| driver 10 | AC Minibus | Good | 109.13 | 102.92 | 106.02 |
| driver 11 | Cab | Good | 104.63 | 139.56 | 122.09 |
| driver 12 | Bus | Good | 122.92 | 119.02 | 120.97 |
| driver 13 | Van | Good | 114.06 | 141.52 | 127.79 |
| driver 14 | Bus | Good | 123.87 | 152.65 | 138.26 |
| driver 15 | Cab | Good | 105.75 | 121.73 | 113.74 |
| driver 16 | Truck | Good | 100.38 | 87.13 | 93.75 |
| driver 17 | AC bus | Very Good | 128.56 | 154.46 | 141.51 |
| driver 18 | Jeep | Satisfactory (Non AC) | 53.29 | 40.17 | 46.729 |
| driver 19 | Van | Good | 121.25 | 129.50 | 125.38 |
| driver 20 | Bus | Good | 123.44 | 147.40 | 135.42 |
| driver 21 | Van | Good | 135.96 | 111.46 | 123.71 |
| driver 22* | 3-wheel | Satisfactory | 34.25 | - | 34.25 |
| driver 23* | 3-wheel | Satisfactory | 72.44 | - | 72.438 |
| driver 24* | Van | Satisfactory | 98.60 | - | 98.604 |
| driver 25* | Van | Satisfactory | 54.33 | - | 54.333 |
| driver 26* | Minibus | Good | 97.17 | - | 97.167 |
| driver 27* | Van | Satisfactory (front AC only) | 89.02 | - | 89.021 |
| driver 28* | Van | Satisfactory (front AC only) | 55.67 | - | 55.667 |

*Newly recruited drivers appointed w. e. f. 15th Feb 2018. (Worked only for 9 months)

From the figures of the above table, it is clear that drivers assigned to vehicles of good condition have a high average overtime rate per month.

By observing overtime hours worked by the driver and the type of vehicle the driver is assigned, it is clear that all the bus drivers have done more OT. It is mainly due to the field trips of students which normally take 2 to 3 days. The Drivers of AC vans have also earned more OT, as an AC van accompanies a bus most of the time to transport staff members who are in charge of the students. A driver who spends the night outside the University while on duty is eligible for 15 hours of overtime per day.

The number of overtime hours worked by each driver for each month of the years 2017 and 2019 and their averages are depicted in the chart below.

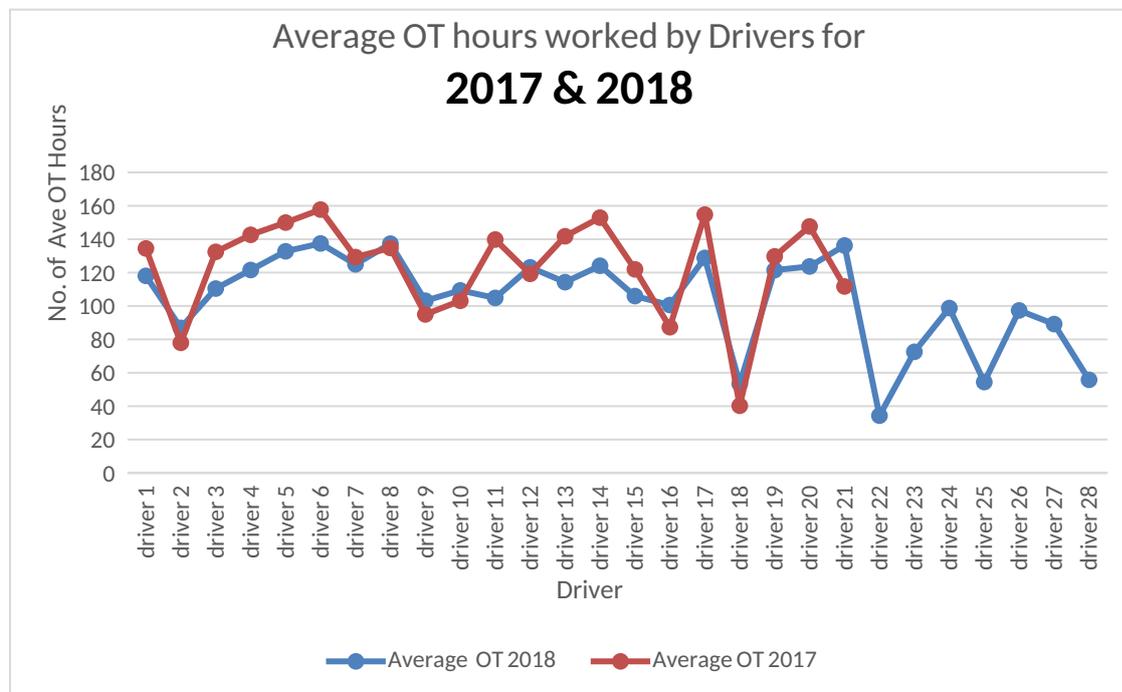

Chart 1

From the above table and chart 1, it is clear that all the bus drivers and drivers assigned to AC vans have earned more overtime than the others. All the drivers who have earned low OT are those assigned to old and less comfortable vehicles.

In order to find whether the performance of drivers has any effect on attracting more trips to them, a group of staff members who frequently travel in the University vehicles was picked and they were requested to rate each driver on a ten-point scale considering the skilfulness, patience, responsibility, customer service and care for the vehicle. The overall performance was calculated by taking the aggregate of the scores.

The average of the ratings received by the drivers is depicted in table 2.

It is clear that except for two drivers, all the others have scored high marks on skillfulness and responsibility. It is noted that one driver (driver 6) has scored full marks for customer service and care for the vehicle and he has the highest average of overtime hours among the drivers.

| Driver | Skillfulness | Patience | Responsibility | Customer service | Care for the vehicle | Overall Performance* |
|---|---|---|---|---|---|---|
| driver 1 | 9 | 9 | 9 | 9 | 9 | 45 |
| driver 2 | 8 | 4 | 5 | 5 | 5 | 27 |
| driver 3 | 9 | 7 | 9 | 7 | 8 | 40 |
| driver 4 | 5 | 6 | 5 | 5 | 5 | 26 |
| driver 5 | 9 | 8 | 8 | 8 | 9 | 42 |
| driver 6 | 9 | 9 | 9 | 10 | 10 | 47 |
| driver 7 | 9 | 8 | 8 | 9 | 8 | 42 |
| driver 8 | 9 | 7 | 8 | 8 | 8 | 40 |
| driver 9 | 7 | 8 | 7 | 5 | 7 | 34 |
| driver 10 | 9 | 9 | 8 | 8 | 8 | 42 |
| driver 11 | 9 | 7 | 7 | 7 | 8 | 38 |
| driver 12 | 8 | 8 | 8 | 8 | 8 | 40 |
| driver 13 | 8 | 8 | 8 | 8 | 8 | 40 |
| driver 14 | 9 | 8 | 8 | 7 | 8 | 40 |
| driver 15 | 9 | 8 | 8 | 6 | 8 | 39 |
| driver 16 | 9 | 7 | 8 | 3 | 4 | 31 |
| driver 17 | 9 | 8 | 8 | 7 | 8 | 40 |
| driver 18 | 9 | 8 | 8 | 7 | 8 | 40 |
| driver 19 | 9 | 8 | 8 | 8 | 8 | 41 |
| driver 20 | 8 | 8 | 8 | 8 | 8 | 40 |
| driver 21 | 9 | 8 | 8 | 8 | 8 | 41 |
| driver 22 | 4 | 7 | 4 | 5 | 4 | 24 |
| driver 23 | 9 | 8 | 8 | 7 | 8 | 40 |
| driver 24 | 8 | 9 | 9 | 9 | 9 | 44 |
| driver 25 | 5 | 5 | 5 | 5 | 4 | 24 |
| driver 26 | 9 | 9 | 8 | 8 | 9 | 43 |
| driver 27 | 9 | 9 | 8 | 9 | 8 | 43 |

| driver 28 | 7 | 6 | 5 | 6 | 5 | 29 |

*Marks given out of 50

Table 2

The average overtime hours worked and the average performance of the drivers are depicted in the chart below.

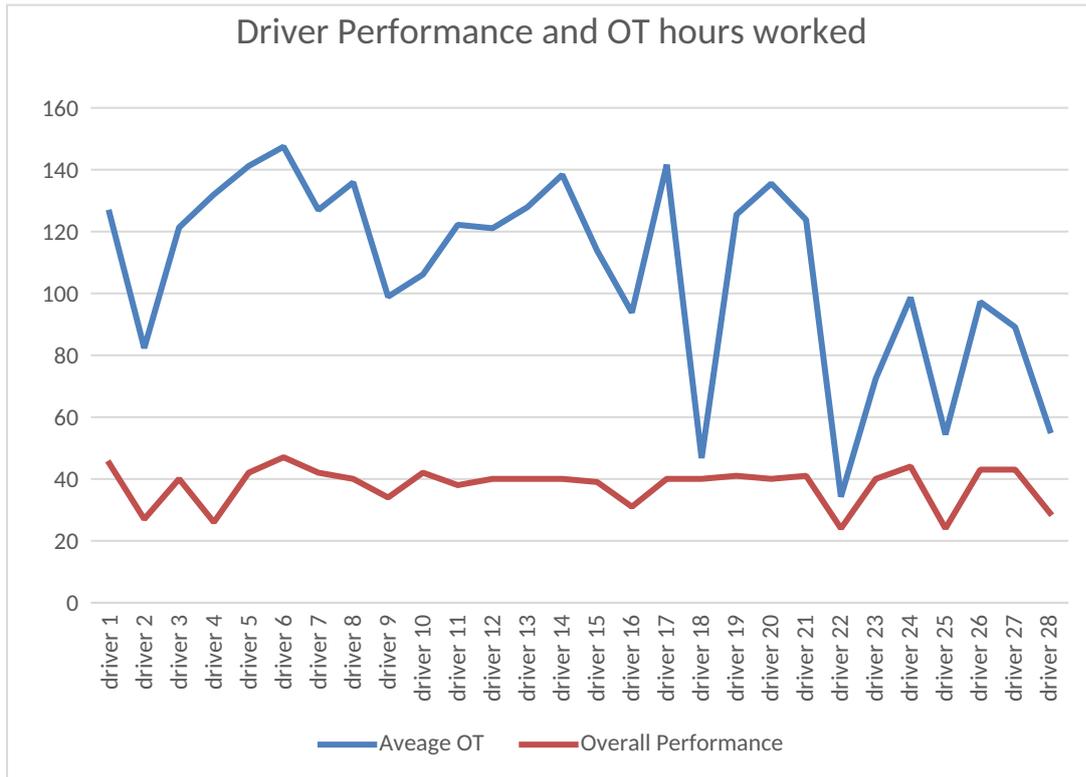

Chart 2

From the above chart, it is apparent that the majority of the drivers performed well and their services are acceptable except for a few drivers who showed a dip in their scores for performance.

It is noted that the drivers who showed a dip in the performance curve also showed a dip in their OT hours curve as well.

The drivers who performed well also showed variation in the number of OT hours worked. The reason could be the condition of the vehicle they are assigned to.

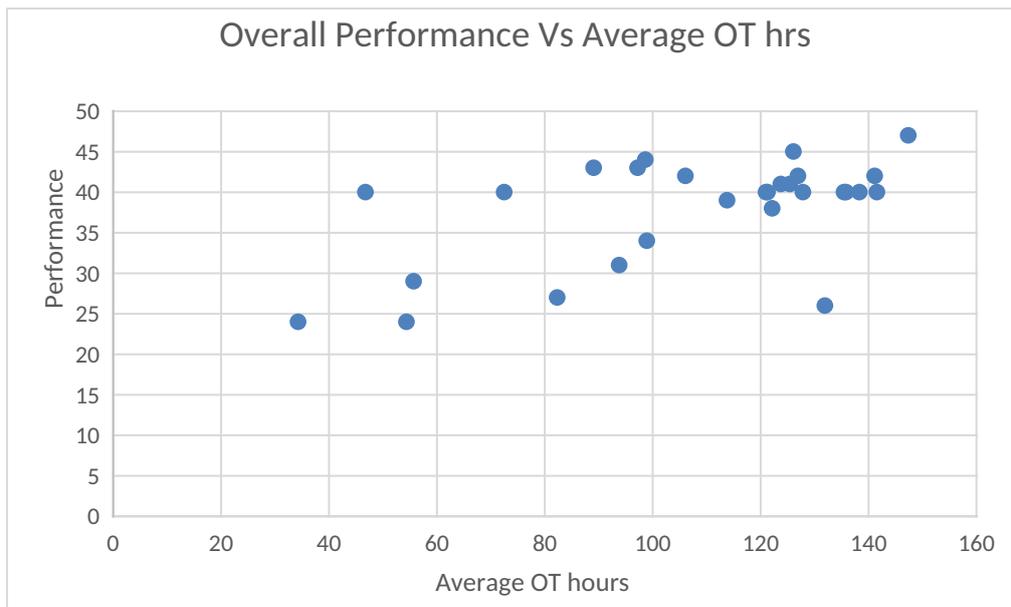

Chart 3

From the above scatter diagram it can be derived that there is no clear linear relationship between the performance of drivers and the number of OT hours worked by them.

**Conclusion**

This study attempted to analyze whether there is a relationship between the performance of drivers and the number of overtime hours worked by them. The drivers assigned to University Officers were excluded from the study as the number of OT hours worked by them directly related to the time of arrival and departure of the Officer, the driver is assigned to. Therefore, 28 drivers in the drivers' pool were subjected to the study.

The number of OT hours worked by the drivers in the pool for the years 2017 and 2018 were extracted from the Overtime Registers and feedback received on the performance of drivers from staff members who frequently travelled in the University vehicles were used for this study. The overall performance of a driver was decided by taking the aggregate of marks received by him for the traits: skilfulness, patience, responsibility, customer service, and care for the vehicle. The type of vehicle the driver is assigned for is also taken into account in the analysis of this study.

 The study revealed that there is no significant relationship between the performance of the drivers and the number of overtime hours worked by them but the type of vehicle and the condition of the vehicle has an effect on attracting long journeys to them which enable them to earn more overtime hours.

The Bus drivers in general have earned a high number of overtime hours especially due to the field trips of students. The drivers who are assigned to AC vans of good condition have also done a higher number of OT. The drivers of old and non AC vehicles have done less Overtime work.

From the above, it can be concluded that the disparity of distribution of overtime hours among the drivers is mainly due to the type and condition of the vehicle and not due to the performance.